# Neutronic analysis on potential accident tolerant fuel-cladding combination $U_3Si_2$-FeCrAl

CHEN Shengli[1]    YUAN Cenxi[1,*]

[1] Sino-French Institute of Nuclear Engineering and Technology, Sun Yat-sen University, Zhuhai, Guangdong 519082, China

[*]Corresponding author. *E-mail address: yuancx@mail.sysu.edu.cn*

## Abstract

Neutronic performance is investigated for a potential accident tolerant fuel (ATF), which consists of $U_3Si_2$ fuel and FeCrAl cladding. In comparison with current $UO_2$-Zr system, FeCrAl has a better oxidation resistance but a larger thermal neutron absorption cross section. $U_3Si_2$ has a higher thermal conductivity and a higher uranium density, which can compensate the reactivity suppressed by FeCrAl. Based on neutronic investigations, a possible $U_3Si_2$-FeCrAl fuel-cladding system is taken into consideration.

Fundamental properties of the suggested fuel-cladding combination are investigated in a fuel assembly. These properties include moderator and fuel temperature coefficients, control rods worth, radial power distribution (in a fuel rod), and different void reactivity coefficients. The present work proves that the new combination has less reactivity variation during its service lifetime. Although compared with the current system, it has a little larger deviation on power distribution, a little less negative temperature coefficient and void reactivity coefficient, and the control rods worth of it is less importance, variations of these parameters are less important during the service lifetime of fuel. Hence, $U_3Si_2$-FeCrAl system is a potential ATF candidate from a neutronic view.

## 1. Introduction

After the Fukushima nuclear disaster in 2011, extensive focuses on the accident tolerant fuel (ATF) have been developed into seeking of advanced nuclear fuel and cladding options. Many materials for fuel and cladding have been investigated over the past years. Three potential approaches have been proposed for the development of the fuel and cladding with enhanced accident tolerance [1]:

1. Modifications of current zircaloy alloy in order to improve the oxidation resistance, including the coating layer design.

2. Replacement of zircaloy cladding by an alternative high-performance oxidation resistant cladding.
3. Improvement or replacement of the ceramic oxide fuel.

The cladding material should have good oxidation resistance, proper delay of the ballooning and burst [2], stress resistance, and small thermal neutron absorption cross section. Among a large number of candidates, stainless steels have better mechanic performance than the current zircaloy-4 alloy. FeCrAl is a potentially promising excellent cladding material [1][3][4]. For example, its oxidation rate is at least two orders of magnitude lower than that of zircaloy [1]. On the other hand, FeCrAl is better to be applied as monolithic cladding than coating, in consideration of matching the thermal expansion coefficient [2], the diametrical compression [2], the volumetric and microstructural evolution [2], the high temperature oxidation protection [5], and the inter diffusion with zirconium [6], which is the reason that the material, Cr coated zircaloy, which is of better mechanic performance has not been selected [7]. In the present work, monolithic FeCrAl is chosen as the potential ATF cladding material.

Many properties have to be considered for the investigation of fuel materials, such as the heavy metal density, the melting point, and the thermal conductivity. Uranium mononitride (UN) based composite fuels may have potential benefits when applied in light water reactor because of its enhanced thermal conductivity and large fuel density. However, UN chemically reacts with water [8], especially at high temperature. Additional shielding material $UN/U_3Si_5$ has been studied to overcome the defect [9]. But a problem still exists, which is the determination of the percentage of $U_3Si_5$ to prevent the fuel from reacting with water in an accident condition. One of the possible solutions is the Tristructural-isotropic (TRISO) fuel design [10][11].

Uranium-silicon binary system, which is thermodynamically stable, is another potential fuel [12]. Among the multiple compounds, $U_3Si$ and $U_3Si_2$ are the best candidates due to their high uranium densities. However, $U_3Si$ swells considerably under irradiation and dissociates into $U_3Si_2$ and solid solution U above 900°C, which is below some possible temperatures in uranium silicide fueled pins [13]. $U_3Si_2$ has promising records under irradiation in research reactor fuels and maintains several advantageous properties compared with $UO_2$.

In consideration of neutronic performance, the thermal neutron absorption cross section of FeCrAl is 2.43 barns, while that of Zircaloy is 0.20 barns [14]. In order to compensate the larger cross section, one needs to decrease the thickness of cladding and/or increase the quantity of fissile nuclides in the fuel. Under economical and safety considerations, the present work chooses nuclear fuel $U_3Si_2$ which has higher uranium density than the current $UO_2$ fuel.

Since the neutronic performance of FeCrAl is different from current Zircaloy and $U_3Si_2$ has different uranium percentage from $UO_2$, analysis on neutronic performance is necessary for the new fuel-cladding system. Neutronic analyses on $U_3Si_2$-FeCrAl, $U_3Si_2$-SiC, and $UO_2$-Zr are performed based on an $I^2$S-LWR [15]. In addition, in order to study the core performance under normal and accident conditions, it is also necessary to investigate the fundamental properties of nuclear reactor core besides the neutron economy analysis, such as moderator and fuel temperature coefficients, control rods worth, radial power distribution, and different void reactivity coefficients.

## 2. Methodology and input parameters

The neutronic behavior is performed using TRITON and KENO-VI modules from SCALE 6.1 [16][17][18]. SCALE provides a "plug-and-play" framework with 89 computational modules, including three Monte Carlo and three deterministic radiation transport solvers. It includes state of the art nuclear data libraries and problem-dependent processing tools for both continuous-energy and multigroup neutronic calculations, multigroup coupled neutron-gamma calculations, as well as activation and decay calculations. TRITION is a module for isotopic depletion calculation. KENO-VI is used for critical calculation. The nuclear data used in our simulation is based on ENDF/B-VII.0 238-group neutron library [19].

Monte Carlo codes are more reliable for radial distribution calculation than the deterministic codes due to the self-shielding. The advantage of Monte Carlo codes is especially evident in such calculation for a new fuel-cladding combination considering the less correction formula existing in deterministic codes. The radial power distribution is calculated by the Monte Carlo based code RMC [20]. RMC is a 3-D Monte Carlo neutron transport code developed by Tsinghua University. The code RMC intends to solve comprehensive problems in reactor, especially the problems on reactor physics. It is able to deal with complex geometry, using continuous-energy pointwise cross sections ENFF/B-VII.0 for different materials and at different temperatures. It can carry out both criticality and burnup calculations, which help to obtain the effective multiplication factor and the isotopic concentrations at different burnup level. Monte Carlo method is also used in fundamental nuclear physics. Nuclear structure problem can be solved with traditional nuclear shell model [21][22] or Monte Carlo shell model [23].

2.1 Geometric parameters and model description

The Westinghouse 17×17 assembly design is used, as shown in Fig. 1. Larger rings placed within the lattice represent the guide tubes and instrumentation tube. Control rods are inserted in the 24 tubes except the center tube for instrumentation. When no control rods or instrument are inserted, these tubes are filled with moderator. All the simulations are based on one assembly unit, except the calculation on radial power distribution. The infinite lattice cell is used to calculate the effective resonant cross section.

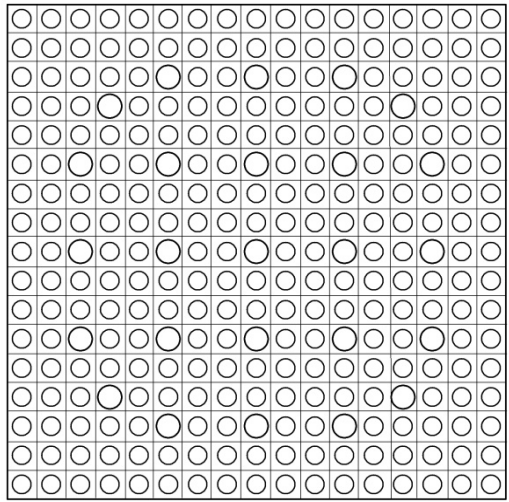

Fig. 1. Westinghouse 17×17 PWR lattice model

In order to maintain the power transfer, pitch-to-rod-diameter (P/D) is kept constant at 1.326. The pellet-cladding gap is also kept constant at 82.55 µm to maintain the thermal conductivity of the gap. The change of cladding thickness is achieved by changing the pellet radius and the inner diameter of cladding. An average value of 630 ppm boron, which is the equivalent concentration at Middle of Cycle (MOC), is placed in the coolant. The other parameters are all presented in Table 1.

| Property | Unit | Value | Reference |
|---|---|---|---|
| Assembly fuel height | cm | 365.76 | [4] |
| Cladding composition | wt% | Zr-4: Fe/Cr/Zr/Sn = 0.15/0.1/98.26/1.49* | |
| | | FeCrAl: Fe/Cr/Al = 75/20/5 | |
| Fuel pellet radius | mm | 4.09575 | |
| Gap thickness | µm | 82.55 | |
| Cladding Inner Radius | mm | 4.1783* | |
| Cladding thickness | µm | 571.5* | |
| Cladding Outer Radius | mm | 4.7498 | |
| Fuel enrichment | % | 4.9* | |
| P/D | | 1.326 | |
| Cladding IR of guide tube | mm | 5.624 | |
| Cladding OR of guide tube | mm | 6.032 | |
| Number of guide tubes | | 25 | |
| Fuel density | g/cm$^3$ | $U_3Si_2$: 11.57 (94.7% theoretical density) | [16] |
| | | $UO_2$: 10.47 (95.5% theoretical density) | [4] |
| Specific power density for reference $UO_2$-Zr | MW/MtU | 38.33 | |
| Coolant density | g/cm$^3$ | 0.7119 | |

| Helium density | g/L | 1.625 (2.0 MPa) | [24] |
|---|---|---|---|
| Cladding density | g/cm$^3$ | Zircaloy-4 cladding: 6.56 | [4] |
| | | FeCrAl cladding: 7.10 | |
| Coolant temperature | K | 580 | |
| Fuel temperature | K | 900 | |
| Cladding and gap temperature | K | 600 | |
| Simulation time | EFPD | 1500 | |
| Boron concentration | ppm | 630 | |
| Boundary conditions | | Reflective | |
| Assembly design | | Westinghouse 17×17 PWR fuel rod | |

\* Values only for the reference case

Table 1. Model specification

In Table 1, the density of helium gas is roughly calculated by the ideal gas equation with a pressure of 2.0 MPa [24]. The density of helium gas has little influence on neutronic analysis. It should be noted that both fuel enrichment and cladding thickness are changed in the following discussions in order to study the sensitivity of the two parameters.

2.2 EOC reactivity calculations

The batch-specific parameters of a typical Westinghouse PWR core are given in Table 2 [25]. The core fraction volume is defined by the number of assemblies of the total 193 assemblies presented in a PWR core for each depletion cycle. The relative assembly power is the ratio between power per batch and the average power of the core. The Effective Full Power Days (EFPDs) achieved at the EOC represent the cycle EFPDs for each batch.

| Batch | Number of assemblies | Core fraction vol% ($V_b$) | Relative assembly power ($P_b$) | EFPDs achieved at EOC ($e_b$) |
|---|---|---|---|---|
| 1 | 73 | 38% | 1.25 | 627 |
| 2 | 68 | 35% | 1.19 | 1221 |
| 3 | 52 | 27% | 0.40 | 1420 |
| Total | 193 | 100% | - | - |

Table 2. Distribution in Population and Power per Fuel Cycle Batch in a Typical Westinghouse PWR

The infinite multiplication factor (k-infinity) at 627, 1221, and 1420 EFPDs are used to estimate the EOC reactivity $\Delta k_{core}$, the difference of reactivity in the reactor core level compared with that of the current PWR core, for each set of fuel geometry. According to the equivalent reactivity method described in Ref. [26], the EOC reactivity for each case is compared with that of a reference case (standard PWR fuel rod containing 4.9% enriched UO$_2$ pellets). The average eigenvalue difference of the core can be estimated by the formula below:

$$\Delta k_{core} = \frac{\Sigma_b \Delta k_{\inf-b}(e_b) P_b V_b}{\Sigma_b P_b V_b}, \qquad (1)$$

where $\Delta k_{\inf-b}$ is the difference of k-infinity between the fuel design under consideration and the reference case for batch b as a function of exposure ($e_b$). The EOC EFPD values in Table 2 are used to quantify the level of exposure for each batch. The power weighting factor ($P_b$) approximates the power distribution in the core to provide the contribution of each batch. The number of assemblies per batch is denoted as $V_b$. The EOC reactivity calculation is a simple but accurate method for the estimation of the cycle length.

Through the reactivity analysis at the EOC, the U$_3$Si$_2$-FeCrAl fuel-cladding combination is suggested in the following section to replace the current combination in the condition of no reduction in the cycle length.

2.3 Moderator temperature coefficient calculation

For a new fuel-cladding combination, the Moderator Temperature Coefficient (MTC) is the safety parameter to be firstly analyzed because it reflects the feedback of reactivity during incident and accident conditions in reactor core. To analyze the MTC, only the temperature of moderator is to be suddenly changed to compare the reactivity at different temperatures. In the present study, the MTC is analyzed with 630 ppm constant boron concentration for feasibility. The density variation with temperature should be considered for liquid phase. The thermal expansion coefficient of the moderator is $\beta_M = 0.00329871/°C$ [27].

2.4 Control rod worth as a function of EFPDs

In the design and the analysis of a nuclear reactor core, the reactivity worth of control rods (i.e. their efficiency in neutron absorption) is another important parameter. The control rods worth is affected by the fuel burnup due to the isotopes variation during operation. For global critical safety consideration, the present work explores the integral worth of traditional and currently used Ag(80%)-In(15%)-Cd(5%) control rods [28] in an assembly as a function of EFPDs because the sub-critical condition must be ensured in the case of full insertion of control rods.

2.5 Radial distribution of power

Many investigations show a non-uniformed radial power distribution in a fuel rod due to the spatial self-shielding. The strong neutron absorption of $^{238}$U at certain energies induces the significant plutonium production, and local power and burnup near the surface of the fuel rod. Therefore, it is of great importance to investigate the rim effect when considering the new fuel-cladding system. The fuel region is divided into 9 rings, while the gap and cladding are located outside, as shown in Fig. 2.

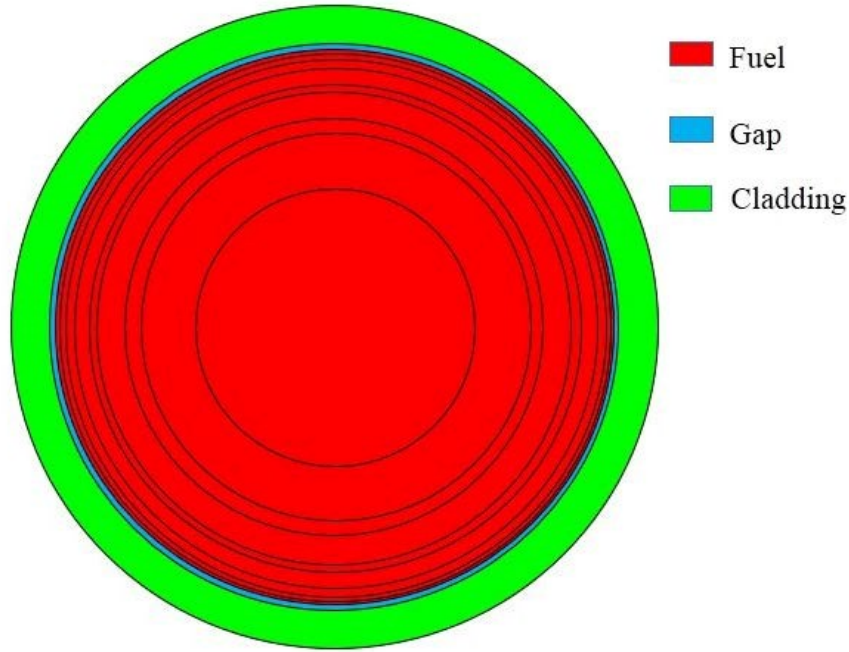

Fig. 2. Radial profile of a fuel-cladding system with fuel region divided into 9 rings

## 3. Results and discussion

3.1 Depletion k-infinity results

Fig. 3 shows the infinite multiplication factor k-infinity of a standard assembly as a function of EFPDs. The reference scenario uses the $UO_2$-Zr fuel rod with parameters shown in Table 1. Results of $U_3Si_2$-FeCrAl fuel rod are also shown with 4.9% enriched $U_3Si_2$ and different cladding thicknesses. For $U_3Si_2$-FeCrAl cases, k-infinity decreases when the cladding thickness increases because of the smaller volume of $U_3Si_2$ and the more neutrons absorbed by FeCrAl.

At the beginning, k-infinity in the reference scenario is larger than that in $U_3Si_2$-FeCrAl cases, which may be caused by the hardened spectrum (presented in section 3.2). The k-infinity of the reference scenario decreases more quickly than in $U_3Si_2$-FeCrAl cases with the increment of EFPDs. One reason is that the burnup in $U_3Si_2$-FeCrAl cases is less than that in the reference case. Another reason is the larger production of the fissile nuclide $^{239}$Pu in $U_3Si_2$-FeCrAl scenarios than that in the reference scenarios, as shown in Fig. 4. The larger concentrations of fissile nuclides give positive contribution to the k-infinity, which makes the k-infinity reduces less quickly in the new system. More $^{239}$Pu are produced because FeCrAl cladding absorbs more thermal neutrons and $^{238}$U nuclei absorb more fast neutrons, especially of which the energy is in the resonance region, to produce $^{239}$Pu at the same EFPDs. Fig. 5 shows that $N_f$ concentration (equivalent $^{235}$U concentration considering fission) of the new combination (with 4.9% enriched $U_3Si_2$-FeCrAl system and the same cycle length as the reference case) varies less than that of the reference scenario. The reason is that more $^{239}$Pu, of which the fission cross section is almost 2.5 times that of $^{235}$U [29], are produced.

Although $U_3Si_2$-FeCrAl system has less variation in reactivity from low to high burnup at constant boron concentration, the boron coefficient is also less negative due to the hardened spectrum, which is also presented in $UO_2$-FeCrAl combination [30] and UN-$U_3Si_2$ combination [31]. Higher boron concentration is needed at the BOC for critical condition, which is presented in an I$^2$S-LWR [15].

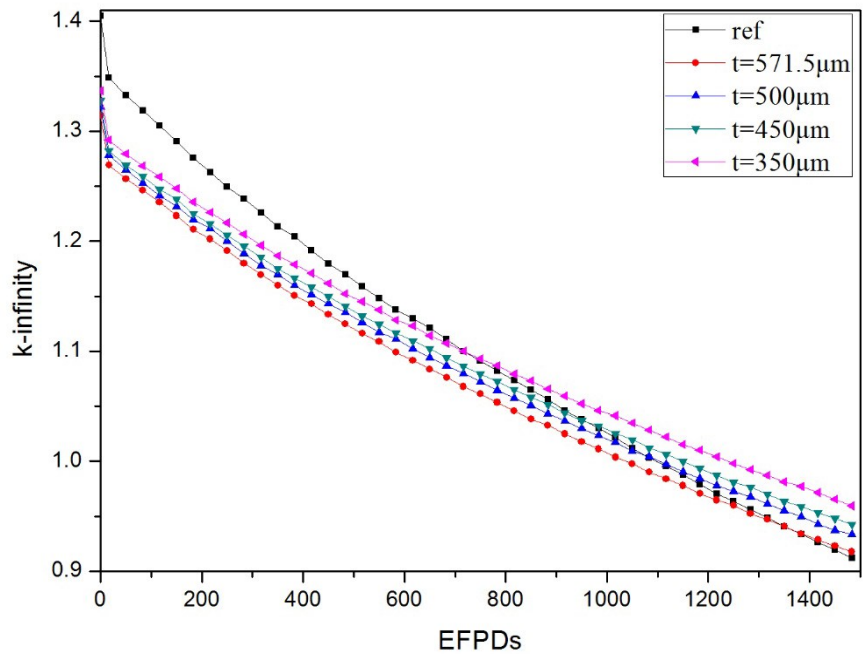

Fig. 3. k-infinity versus EFPDs for 4.9% enriched fuel

*Remark: in all figures, ref represents the reference case and new represents the 4.9% enriched $U_3Si_2$ fuel and 450 μm thickness of FeCrAl cladding.*

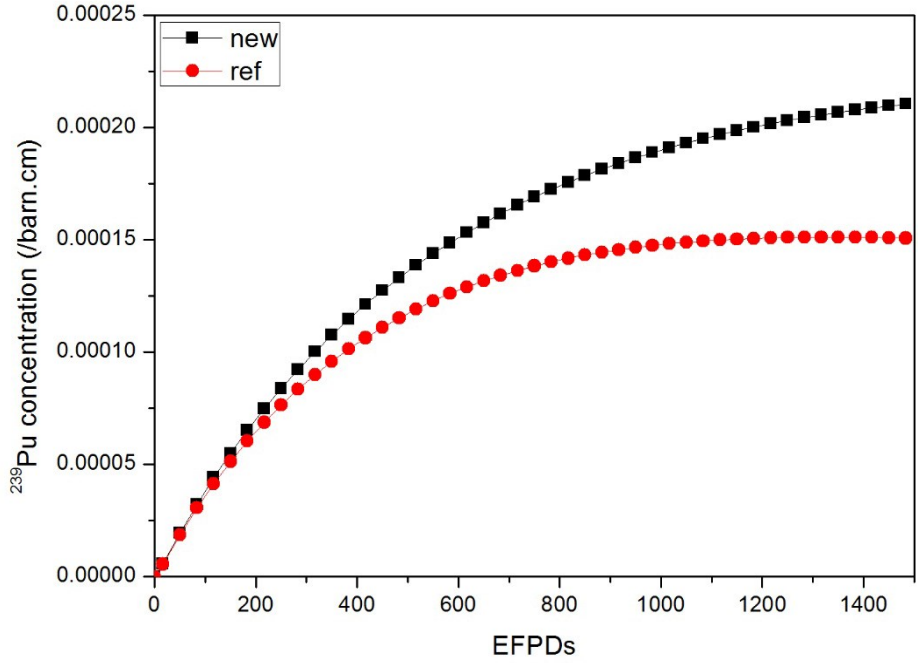

Fig. 4. Concentration of $^{239}$Pu versus EFPDs

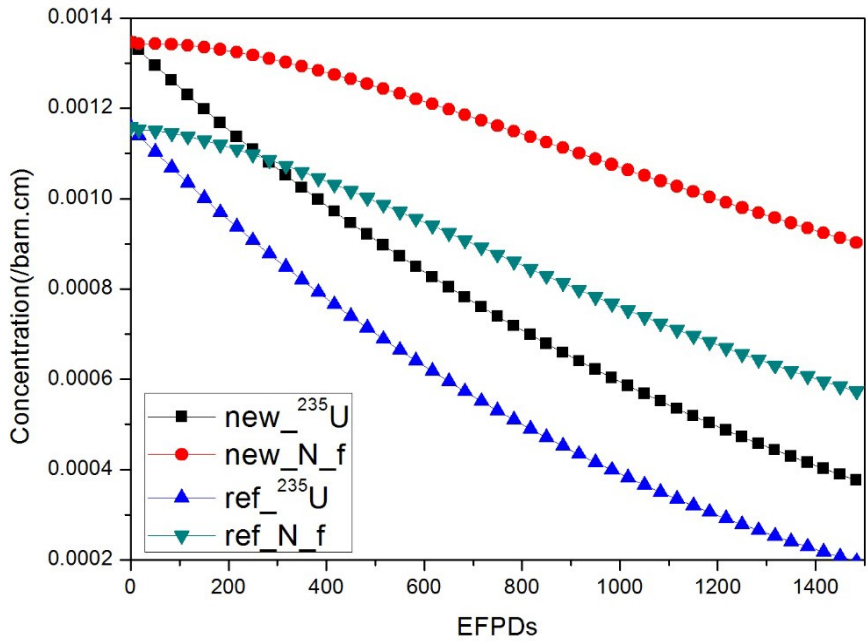

Fig. 5. Concentration of $^{235}$U and equivalent fissile isotopic versus EFPDs

Table 3 displays the $\Delta k_{core}$ corresponding to each cladding thickness and uranium enrichment. In this work, one of the most important objectives is to find the relationship between the uranium enrichment and the cladding thickness without reduction in cycle length. It is thus not necessary to analyze the reactivity of combinations with $\Delta k_{core}$ far away from zero. The specific power per metric tonne of uranium (MW/MtU) corresponds to the constant power of 18.0 MW/assembly, which means 38.33 MW/MtU specific power density for reference 4.9% enriched $UO_2$-Zr case. The positive (negative) values of $\Delta k_{core}$ signify longer (shorter) cycle length than the reference scenario.

| $^{235}U$ enrichment e (%) | Cladding thickness t (µm) | | | | | |
|---|---|---|---|---|---|---|
| | 571.5 | 540 | 500 | 450 | 400 | 350 |
| 4.5 | - | - | - | - | - | -0.00484 |
| 4.6 | - | - | - | -0.01465 | -0.00669 | 0.00078 |
| 4.7 | -0.03060 | -0.02456 | -0.01778 | -0.00935 | -0.00213 | 0.00522 |
| 4.8 | -0.02499 | -0.01933 | -0.01304 | -0.00386 | 0.00377 | 0.01097 |
| 4.9 | -0.02007 | -0.01490 | -0.00806 | 0.00019 | 0.00827 | 0.01565 |
| 5.0 | -0.01460 | -0.00929 | -0.00238 | 0.00528 | 0.01242 | 0.02018 |
| 5.1 | -0.00968 | -0.00456 | 0.00156 | 0.00999 | 0.01748 | - |
| 5.2 | -0.00452 | 0.00001 | 0.00712 | 0.01538 | - | - |

Table 3. Cycle reactivity difference $\Delta k_{core}$ between the $U_3Si_2$-FeCrAl and reference system

With these values, by linear fitting of $\Delta k_{core}$ as function of fuel enrichment and cladding thickness, one can obtain:

$$\Delta k_{core} = ae + bt + c, \qquad (2)$$

where $e$ and $t$ represent uranium enrichment (%) and cladding thickness (µm), respectively, $a = 0.04980 \pm 0.00042\%$, $b = -1.59488 \pm 0.01050 \times 10^{-4} \mu m$ and $c = -0.17234 \pm 0.00190$ with the coefficient of determination $R^2 = 0.99875$. Positive value of the coefficient $a$ and negative value of the coefficient $b$ are logical because the increment of uranium enrichment increases the cycle length, while the increment of cladding thickness has a converse effect. The coefficient $c$ is negative because when the uranium enrichment is too low, the cycle length in the reference case can never be achieved.

Based on the above results, it can be concluded that 1% uranium enrichment contributes +4980 pcm to $\Delta k_{core}$ at the EOC, while 1 µm more in cladding thickness induces -16 pcm. In other words, a 1% increment in uranium enrichment can compensate the negative reactivity induced by an increment of 312 µm in the thickness FeCrAl cladding.

According to equation (2), the cladding thickness needs to be 450µm to keep the same cycle length and uranium enrichment as in the reference case, 4.9%. The 4.9% enriched $U_3Si_2$ fuel and 450µm thickness of FeCrAl cladding is the new fuel-cladding combination which will be discussed hereafter.

In general, for U$_3$Si$_2$-FeCrAl fuel-cladding system, in order to achieve the same cycle length as the current 4.9% enriched UO$_2$-Zr system, the relationship between uranium enrichment and cladding thickness is taken as:

$$e = -\frac{bt+c}{a} \text{ or } e = 3.2 \times 10^{-3}t + 3.5, \quad (3)$$

In practice, the uranium enrichment and cladding thickness for U$_3$Si$_2$-FeCrAl system should be located in the white area in Fig. 6 to achieve no shorter cycle length compared with the current UO$_2$-Zr system.

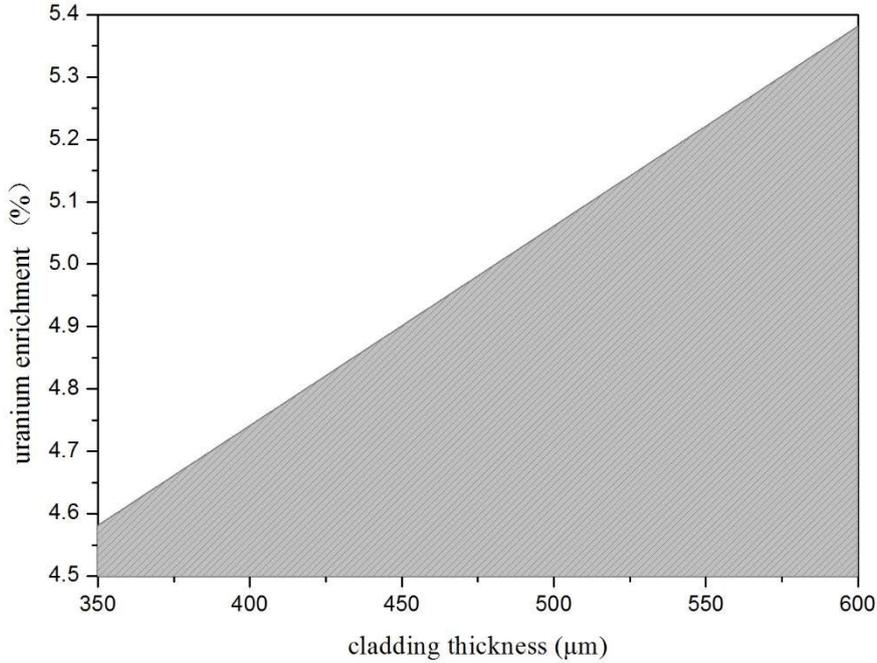

Fig. 6. Feasible combination in consideration of cycle length

Stainless steel of a cladding thickness of 350 μm was used as nuclear fuel cladding [32]. For the same FeCrAl cladding with a thickness of 350 μm, only 4.58% enriched U$_3$Si$_2$ fuel is needed to achieve the cycle length in the reference scenario, while 5.06% uranium enrichment of UO$_2$ fuel is needed, as mentioned in Table 6 in Ref. [3]. It costs less for the fabrication of enriched nuclear fuel due to the lower uranium enrichment.

3.2 Spectral hardening

Spectral hardening is a phenomenon of FeCrAl cladding because of its higher thermal neutron absorption cross section than that of the zirconium alloy. Another reason is that the moderator-to-fuel ratio is smaller in the new combination, which reduces the resonance escape probability and the percentage of thermal neutrons.

The hardening of the neutron spectrum is also found in the transition from low burnup to high burnup. This phenomenon is due to the accumulation of fission products and actinides. A hardened neutron spectrum includes less thermal neutrons, which mainly induce the fission. As a consequence, the reactivity is reduced in the case of spectral hardening. A possible solution to compensate such effect is the increment of the moderator-to-fuel ratio.

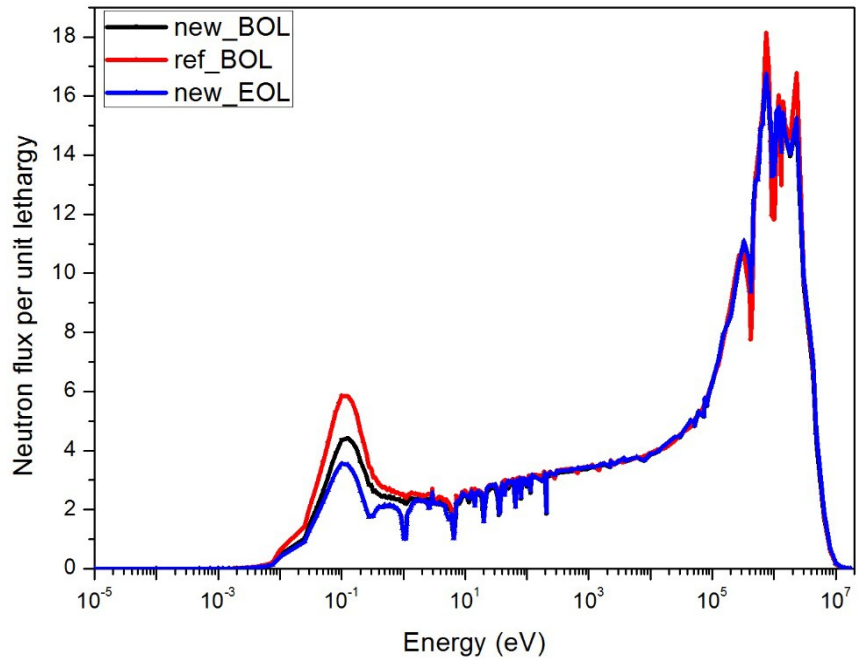

Fig. 7. Spectrum of neutron flux

3.3 Temperature coefficients

The MTC and the Fuel Temperature Coefficient (FTC, also known as Doppler coefficient) are two of the most important temperature feedback parameters in nuclear safety because they automatically control the reactivity of the core when temperature changes. It is thus of great interest to study the two parameters for a new fuel-cladding combination.

From the definition of the reactivity $\rho = \frac{k-1}{k}$, the difference of two reactivity can be calculated using the corresponding multiplication factor $k$ as:

$$\Delta \rho = \rho_2 - \rho_1 = \frac{k_2 - k_1}{k_1 k_2}, \quad (4)$$

In the calculation of temperature coefficients, all concentrations of nuclides in the fuel are extracted at each burnup depth in normal condition. The moderator temperature and the corresponding density (fuel temperature respectively) are changed through the definition of the MTC (FTC respectively). The corresponding multiplication factors are also calculated using the concentrations of nuclides extracted in normal condition calculation. The difference of reactivity can be obtained through Eq. (4). The temperature coefficients are determined by dividing the difference of reactivity by the corresponding difference of temperature.

The MTC is shown in Fig. 8 for the new combination and the reference scenario. The MTC values are determined by the reactivity calculation in normal condition and at 610 K moderator temperature. At the beginning, the MTC value is more negative for new combination due to the hardened spectrum, which reduces the number of thermal neutrons and enhances the importance of the moderator.

After the beginning of the operation, the MTC values become more and more negative. The reason is that the increment of $^{238}$U to $N_f$ ratio (as shown in Fig. 9) enhances the resonance absorption of $^{238}$U, which leads to the decrement of resonance escape probability. Such effect emphasizes the importance of moderator and results in enlarged negative MTC values with burnup. The less negative MTC value of the new combination can also be explained by the lower $^{238}$U to $N_f$ ratio (as shown in Fig. 9) due to the hardened spectrum with FeCrAl cladding, as explained in section 3.1.

It should be remarked that the MTC is calculated at 630 ppm boron concentration. In practice, the MTC is less negative at the BOC because of the actually higher boron concentration, and more negative at EOC due to the lower boron concentration.

As shown in Fig. 10, the FTC is always negative because the increment of fuel temperature leads to the broadening of resonance absorption of $^{238}$U. The FTC values become more and more negative with burnup due to the increment of $^{238}$U to $N_f$ ratio, which reflects the impact of $^{238}$U. Moreover, the less negative FTC values of the new combination are due to the smaller $^{238}$U to $N_f$ ratio.

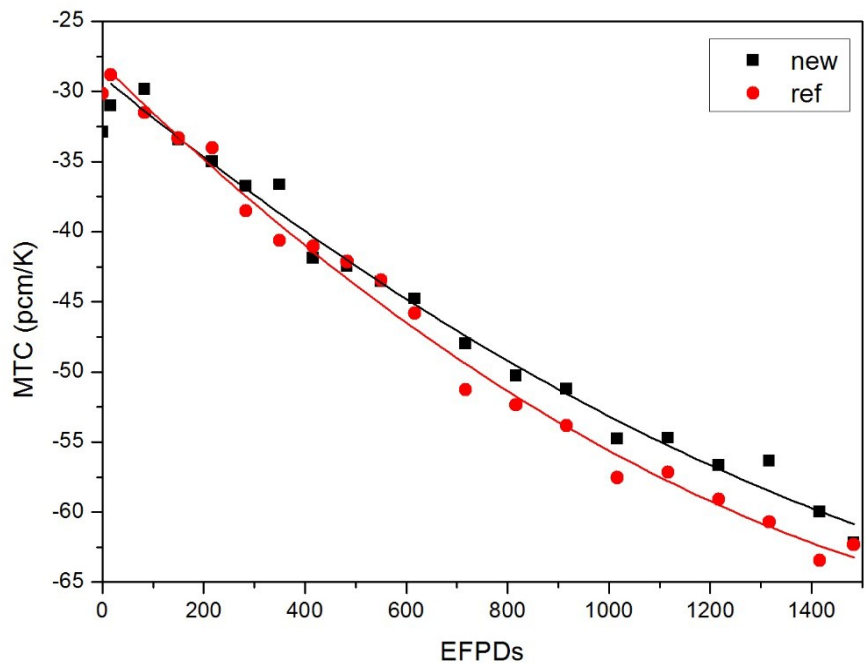

Fig. 8. MTC vs EFPDs

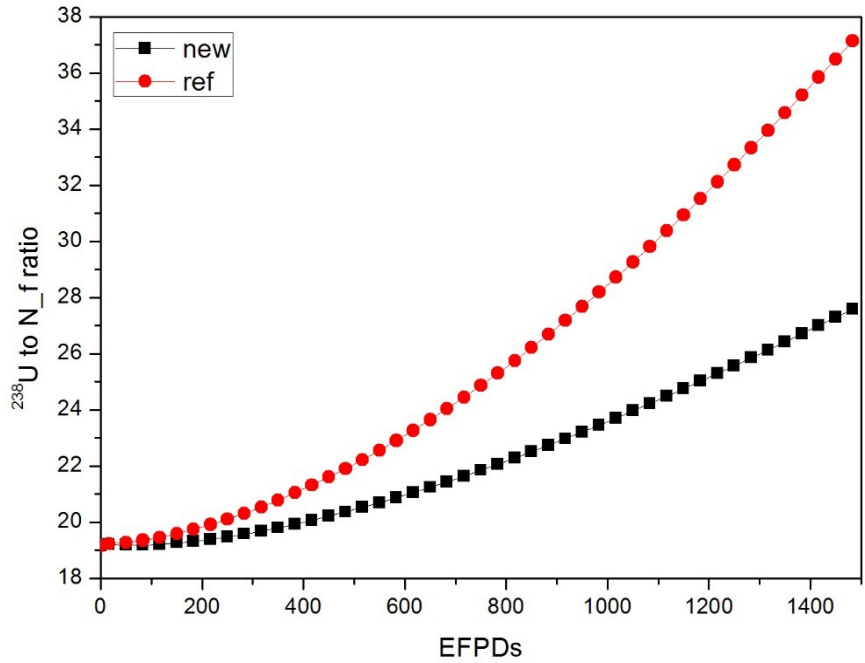

Fig. 9. $^{238}$U to $N_f$ ratio for 630 ppm boron concentration

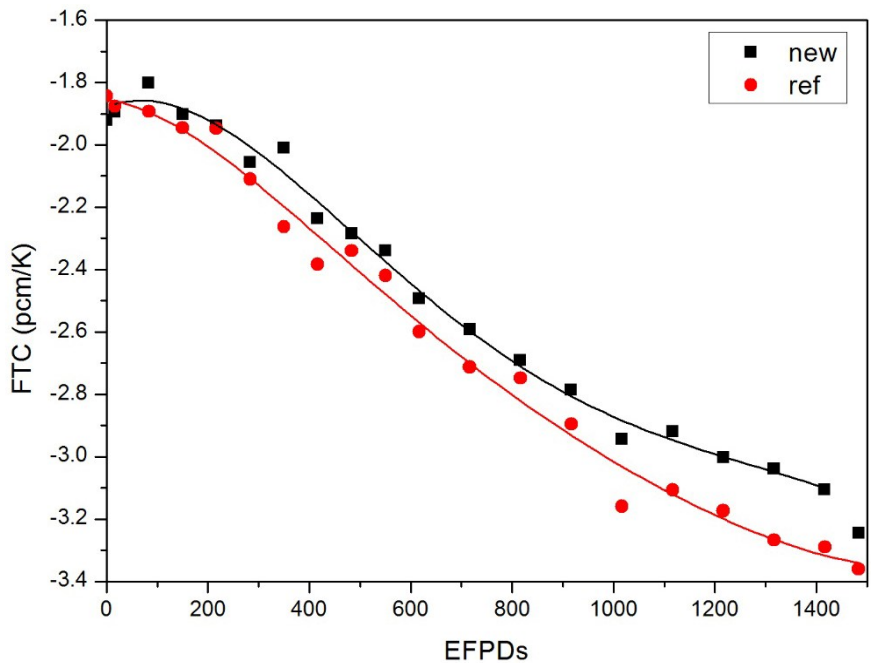

Fig. 10. FTC vs EFPDs

3.4 Moderator temperature sensitivity

The reactivity of a reactor core depends on the moderator temperature. The MTC parameter represents the reactivity induced by sudden change of moderator temperature. It is also important to study the Moderator Temperature Sensitivity (MTS) at different moderator temperatures, especially for a new fuel-cladding combination. The MTS represents the reactivity sensitivity induced by different moderator temperatures design, which is different from current moderator temperature (580 K) in core design. The moderator temperature keeps constant during a cycle length for the MTS calculations because the MTS reflects the reactivity change in normal operation with different temperatures design.

The MTS increases with boron concentration as shown in Fig. 11. The reason is that the decrement of moderator density also reduces the boron density, which has an opposite effect.

In the range of 0 to 500EFPDs, MTS decreases with EFPDs for the same reason as MTC because there is no great difference between different scenarios with the same boron concentration.

Nonetheless, in the range of 500 to 1500 EFPDs, the MTS increases with burnup because of the larger accumulated number of $^{239}$Pu in less reactivity cases. This is caused by the higher moderator temperature during operation, as shown in Fig. 12. Accumulation of $^{239}$Pu has an opposite effect on the increment of moderator temperature due to the large fission cross section.

When burnup is larger than 900 EFPDs, the MTS of the new combination is more negative. This is due to lower burnup than the reference scenario caused by higher uranium quantity in $U_3Si_2$. The MTS value is about -20 pcm/K at 1500 EFPDs for the new combination. In spite of less thermal efficiency with lower moderator temperature, the decrement of moderator temperature is considerable for the new combination in consideration of the prolongation of the service life of fuel, while it is much less effective for the reference scenario. On the contrary, the increment of moderator temperature in the new design largely reduces the fuel life.

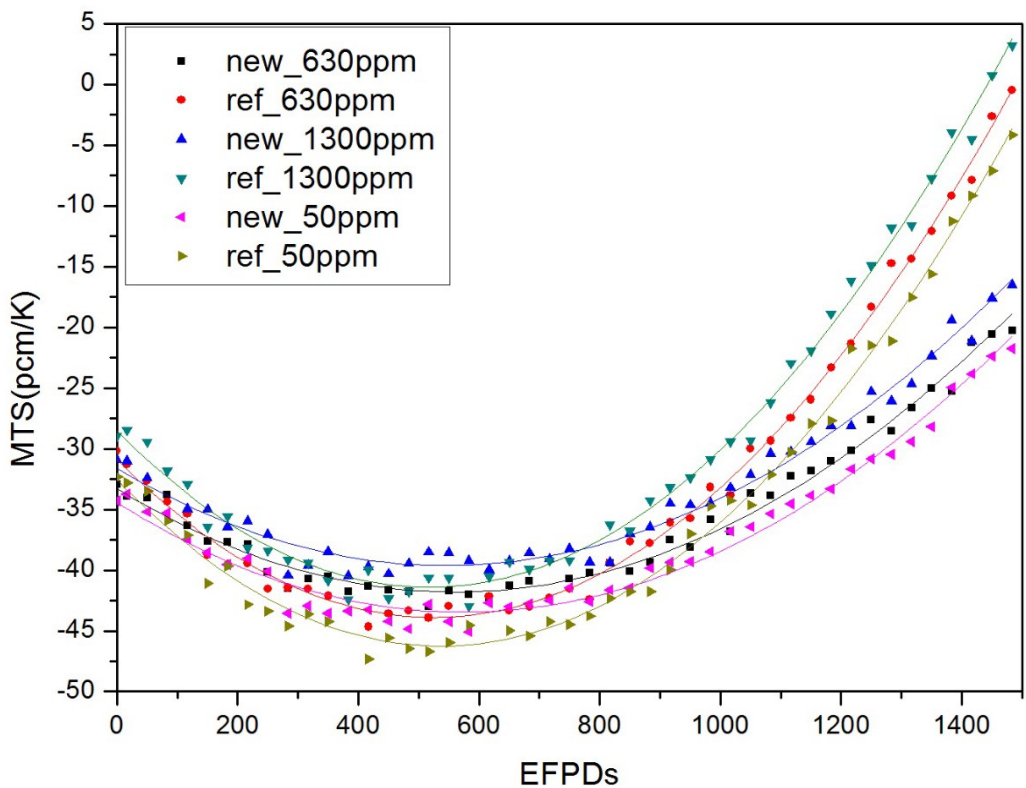

Fig. 11. MTS vs EFPDs

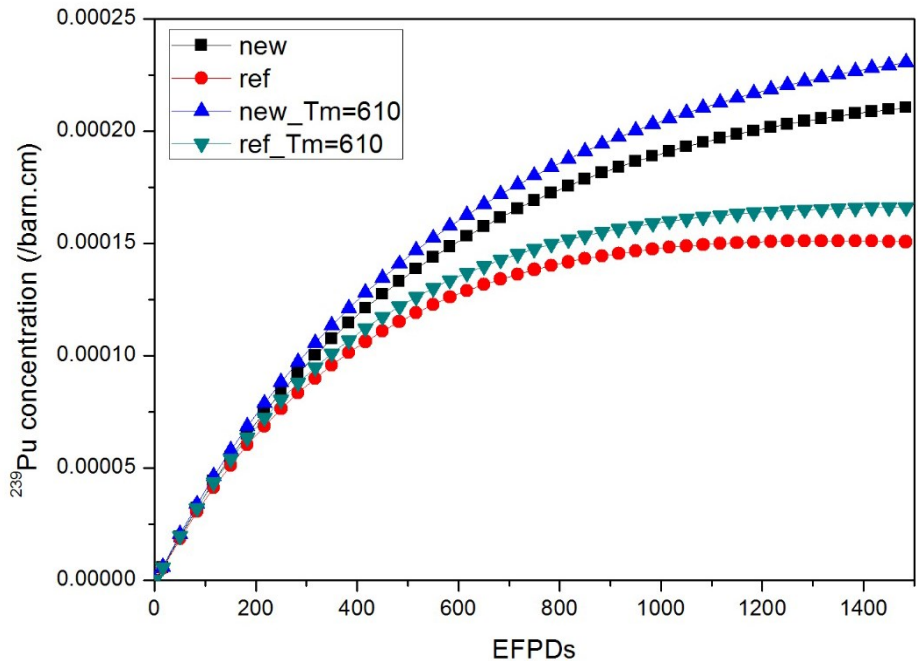

Fig. 12. Concentration of $^{239}$Pu for 580 and 610 K moderator temperature

3.5 Control rods worth

In a similar method to that in the determination of temperature coefficients, calculations are performed to obtain the control rods worth. The multiplication factor is firstly calculated as a function of burnup. The concentrations of all nuclides in the fuel are extracted for each burnup depth. The multiplication factor is obtained at each burnup through the same concentrations and the insertion of all the control rods in 24 tubes. The control rods worth is calculated directly using Eq. (4). It is remarkable that the control rods worth is often expressed by its absolute value.

The integral control rods worth in an assembly is shown in Fig. 13. Control rods worth increases with burnup due to the depletion of $^{235}$U and the decrement of total fissile nuclides. In other words, the ratio of absorbent isotopes in control rods to fissile isotopes in fuel increases with burnup.

In the case of new combination, the control rods worth is smaller than in the reference case. One reason is the higher uranium quantity in the new combination as the control rods are the same. Another reason is that FeCrAl cladding has larger macroscopic thermal neutron absorption cross section, which also plays a similar role as control rods. In other words, spectrum hardening reduces the effect of control rods. It will be of great importance to verify that the control rods worth meets the safety requirement for the new fuel-cladding system.

Lower burnup and lower moderator-to-fuel ratio lead to less variation of the control rods worth for the new combination than in the reference case, which is another advantage of the new fuel-cladding combination.

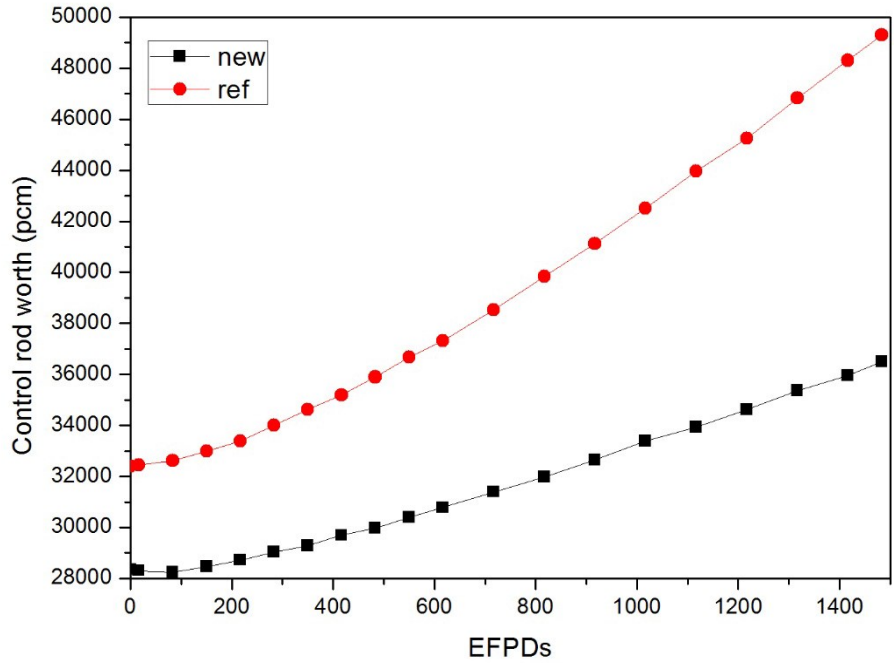

Fig. 13. Control rod worth vs EFPDs

3.6 Radial power distribution

The normalized power distribution versus relative radius is shown in Fig. 14. Because of the spatial self-shielding, some resonance neutrons are shielded from the center of the fuel due to the absorption by the outer fuel. The relative fission power is largest near the surface of the fuel rod and smallest at the center of the pellet. At the BOL, although there is certain difference in neutron absorption ability between two cladding materials, there is no significant difference in the relative power in a fuel rod. During the operation, more plutonium is produced near the surface of the fuel pellet than at the center because of the spatial self-shielding. Additional fissile nuclides cause a sharp increment in fission reactions near the surface of the fuel rod. As shown in Fig. 14, fission profiles in inner regions are flat, while a sharp increment in power is found close to bound. The new combination has a little larger relative power near the surface because more $^{239}$Pu are produced in the outer ring, as shown in Fig. 15.

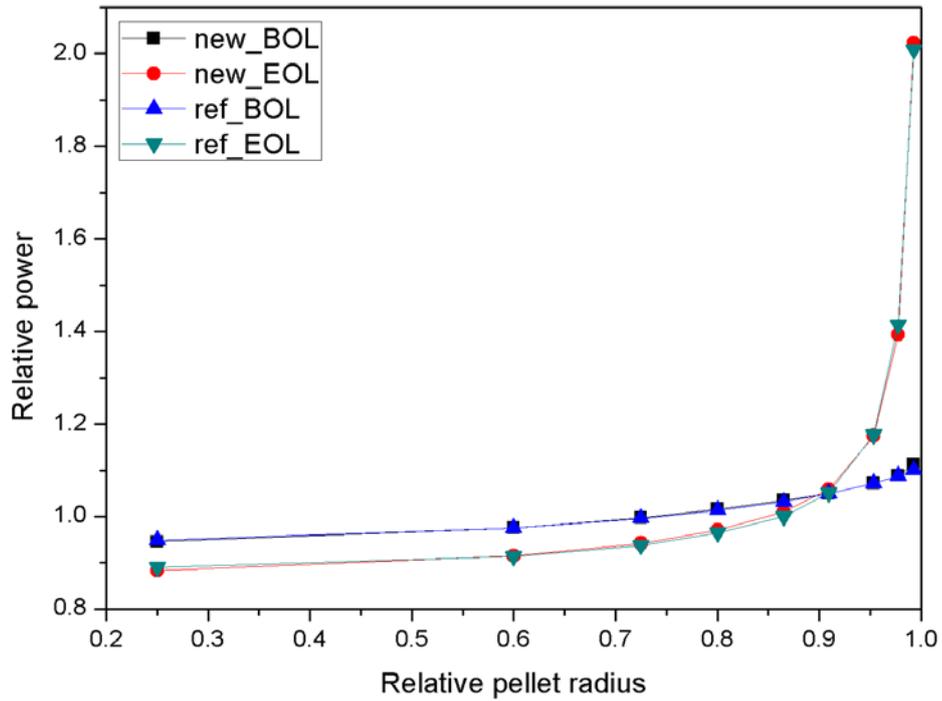

Fig. 14. Relative radial power distribution

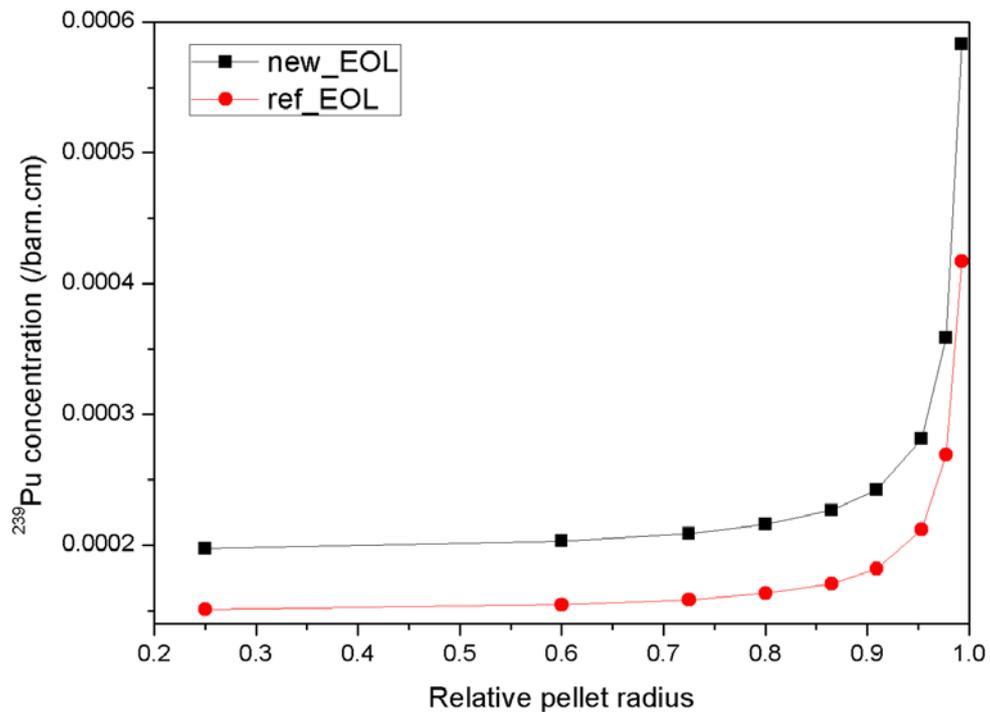

Fig. 15. Radial distribution of $^{239}$Pu

Fig. 15 shows the concentration of $^{239}$Pu at the EOL. $^{239}$Pu is chosen because it is the most important fissile nuclide besides $^{235}$U. More $^{239}$Pu are accumulated in the new combination, which corresponds to the results in Fig. 5. Such fact further emphasizes that cladding materials with higher absorbing ability induce less reactivity in early life due to the hardened neutron spectrum but larger reactivity near the EOL due to the more considerable accumulation of plutonium.

3.7 Void reactivity coefficient

The Void Reactivity Coefficient (VRC) is calculated in a similar way to the calculation of the temperature coefficients and the control rods worth, with the change of the moderator void.

The VRC calculation is of great interest for a new fuel-cladding combination because it describes the variation of reactivity of a reactor in Loss of Coolant Accident (LOCA) scenario. Similar to the MTC, a more negative VRC value is better in the consideration of nuclear safety. It is remarkable that the shape of the VRC as function of EFPDS is similar to that of MTC because the increment of the moderator void has a similar effect to that of the increment of the moderator temperature, which reduces the moderator density. The reason that the VRC is less negative for the new combination is thus the same as that explained in MTC analysis.

At the very beginning, the VRC value increases with burnup. This is not caused by the softened neutron spectrum (it is actually hardened as shown in Fig. 17) but by the abrupt presence of $^{239}$Pu, which has a positive contribution to the VRC [33].

Because of the less negative value of the VRC, the feedback of anti-reactivity in LOCA scenario is less important for the new combination. Therefore, it is another parameter which must be considered in a new core design.

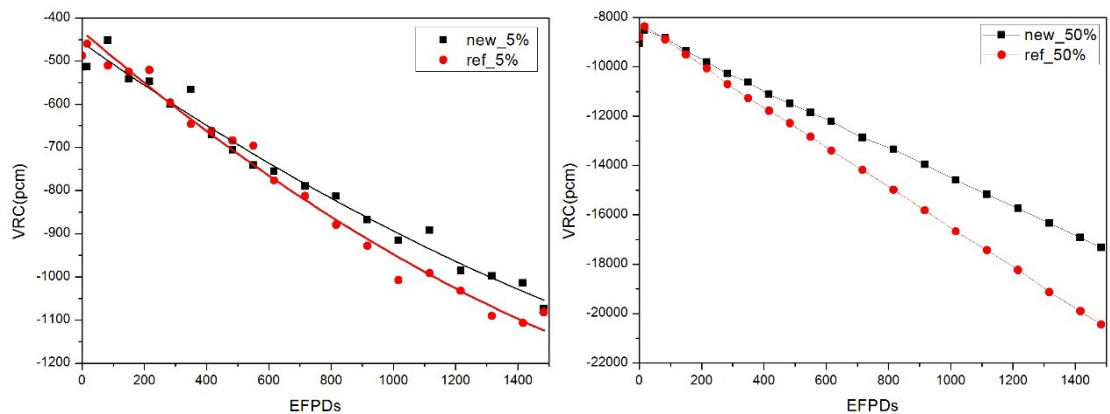

Fig. 16. VRC for 5% and 50% void

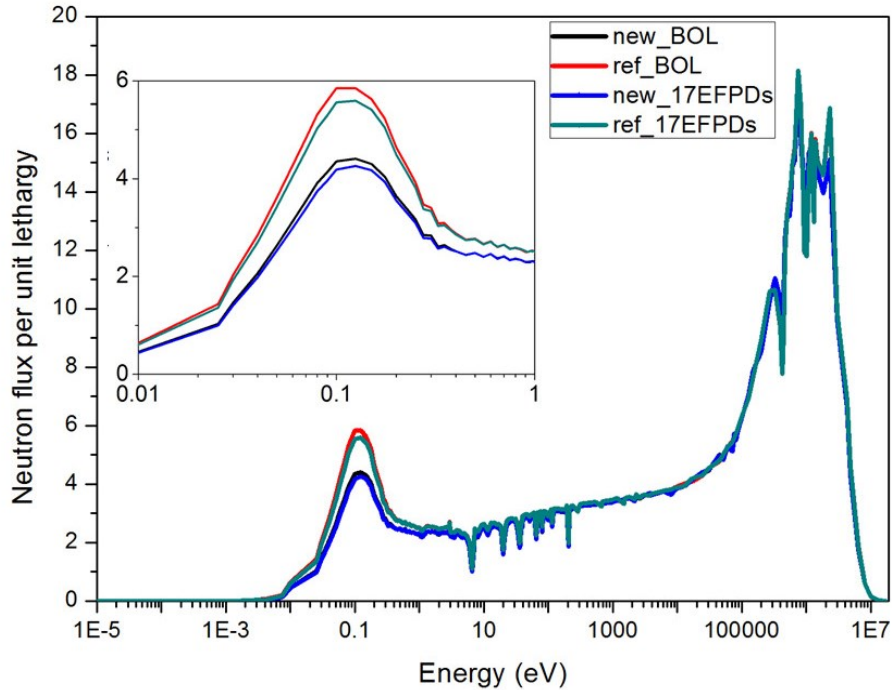

Fig. 17. Neutron flux spectrum

3.8 Void reactivity sensitivity

The VRC is a parameter that represents the reactivity change caused by sudden change of moderator quantity. In the primary circuit of a PWR, a little void in moderator may exist. The increment of the percentage of void in moderator is also used to study the neutronic performance of decreased moderator-to-fuel ratio when the geometry is kept the same as the current core design. Similar to the study of MTS, the Void Reactivity Sensitivity (VRS) is investigated by setting different void percentages in moderator but keeping them constant during the cycle length.

As shown in Fig. 18, the shapes of the VRS as function of EFPDs are similar to that of the MTS. The reason is that the VRS and the MTS are both parameters that measure the reactivity change by modifying the moderator density. The increment of VRS at high burnup is from the large number of accumulation of $^{239}$Pu (as shown in Fig. 19) due to the hardened spectrum. The dependence of k-infinity on the percentage of void of moderator is shown in Fig. 20.

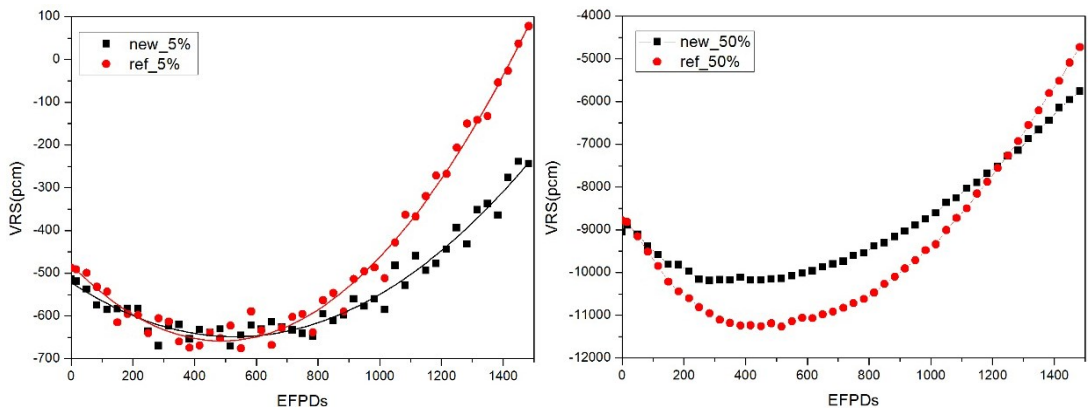

Fig. 18. VRS for 5% and 50% void

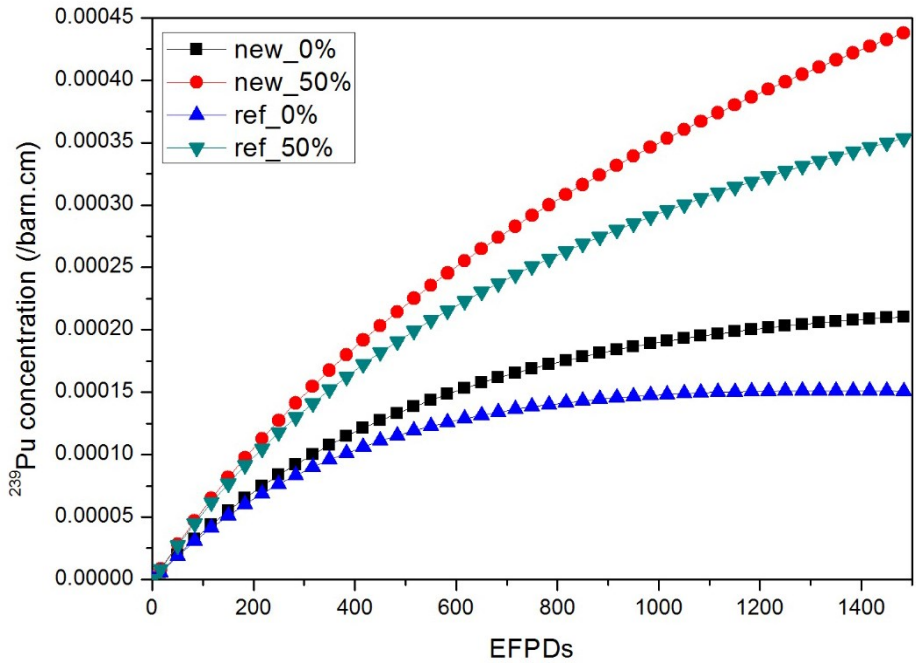

Fig. 19. Concentration of $^{239}$Pu with different percentage of void of moderator

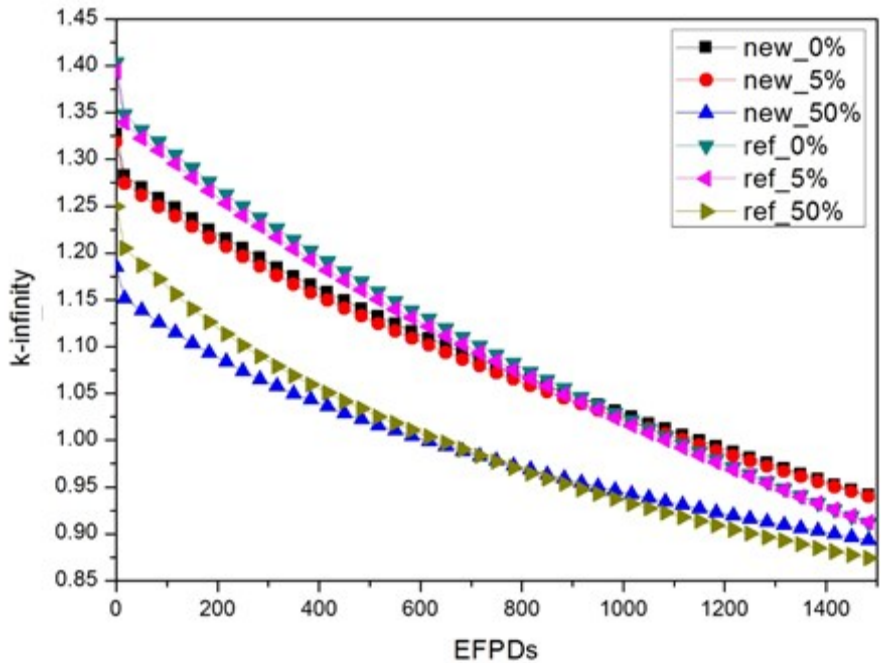

Fig. 20. k-infinity with different percentage of void of moderator

## 4 Conclusion

FeCrAl is a potential cladding material due to its excellent oxidation resistance. But FeCrAl has the neutronic penalty compared with Zr-4 due to its larger thermal neutron absorption cross section. Better neutron economy performance is found for

$U_3Si_2$ fuel than $UO_2$ fuel, which provides a potential solution for all alternative claddings with larger thermal neutron absorption cross section than zirconium alloys. The relationship between uranium enrichment and cladding thickness of $U_3Si_2$-FeCrAl fuel-cladding combination is found by reactivity analysis in the condition of no reduction in the cycle length compared with that of $UO_2$-Zr system. The critical cladding thickness is determined by keeping the uranium enrichment the same as in the reference scenario.

The investigation proves that the new combination has less reactivity variation during its lifetime. The new combination provides less negative feedback for both fuel and moderator temperature reactivity than the reference case. Other disadvantages of the new combination are that it has a little larger deviation in power distribution, a less important control rods worth, and a less negative VRC than the current system. Nonetheless, the MTC, the FTC, the control rods worth and the VRC of the new combination are better in practice in consideration of the stability during the lifetime. Furthermore, it is possible to prolong the service time of the fuel by reducing the moderator temperature for the new combination. It should be noted that the sensitive and uncertainty analyses are also very important for a model calculation, such as the uncertainty analysis on a simple nuclear mass model which has been performed recently [34].

$U_3Si_2$-FeCrAl system is a potential ATF candidate with many advantages compared with potential $UO_2$-FeCrAl system. For example, lower uranium enrichment is required and larger cladding thickness is permitted in this system. Moreover, the thermophysical properties of $U_3Si_2$ have been studied up to a temperature of 1773 K [35], which provides a complementary knowledge for this fuel material. The present work is also one of the possible new potential ATF fuel-cladding combinations proposed by the U.S. DOE NE Advanced Fuels Campaign [36], which shown that the $U_3Si_2$ and stainless steel can be considered as potential fuel and cladding, respectively.

Nevertheless, chemical reactions between water and $U_3Si_2$-FeCrAl system should be taken into the consideration in a reactor with water as the moderator. The possible chemical reaction between $U_3Si_2$ and Al [37] should also be taken into account. Moreover, attention should be paid to different fuel densities due to the different porosities caused during the fabrication of the fuel.

## Conflicts of Interest

The authors declare that there is no conflict of interest regarding the publication of this manuscript.

## Acknowledgments

The authors acknowledge the authorized usage of the RMC code from Tsinghua University for this study. The authors acknowledge also the useful discussions with

Dr. XIAO Sicong. This work has been supported by the National Natural Science Foundation of China under Grant No. 11305272, the Special Program for Applied Research on Super Computation of the NSFC Guangdong Joint Fund (the second phase), the Specialized Research Fund for the Doctoral Program of Higher Education under Grant No. 20130171120014, the Fundamental Research Funds for the Central Universities under Grant No. 14lgpy29, the Guangdong Natural Science Foundation under Grant No. 2014A030313217, and the Pearl River S&T Nova Program of Guangzhou under Grant No. 201506010060.